\documentclass[reqno]{amsart}

\usepackage{amsmath,amsfonts}
\usepackage{epsfig}
\usepackage{graphicx}

\def\epspdf{eps}

\def\alg{\mathfrak{A}}
\def\amp{{\rm Amp}}
\def\bHle{\overline{H}_{\omega,\lambda}}
\def\bn{\bar{n}}
\def\bomzl{\overline{\rho}_0}
\def\bomzlt{\overline{\rho}_t}
\def\brho{\bar{\rho}}

\def\cNum{\mathcal{N}}

\def\cS{\mathcal{S}}

\def\e{\epsilon}

\def\Exp{\mathbb{E}}
\def\Fo{\mathfrak{F}}
\def\Hle{H_{\omega,\lambda}}
\def\Hom{H_{\omega}}
\def\homega{\widehat\omega}
\def\hn{\widehat{n}}

\def\kker{h}
\def\LL{\Lambda_L}

\def\ontop{\genfrac{}{}{0pt}{}}
\def\omzl{\rho_0}

\def\omzlt{\rho_t}
\def\omzlT{\rho_{T/\eta^2}}

\def\sign{{\rm sign}}

\def\talpha{\widetilde{\alpha}}
\def\tf{\widetilde{f}}
\def\tn{\widetilde{n}}
\def\Tor{\mathbb{T}}
\def\Tr{{\rm Tr}}

\def\tk{\widetilde{k}}

\def\Vom{V_\omega}

\def\det{{\rm det}}

\def\C{{\mathbb C}}
\def\N{{\mathbb N}}

\def\R{{\mathbb R}}
\def\Tor{\mathbb T}
\def\Z{{\mathbb Z}}

\def\1{{\bf 1}}

\def\eqnn{\begin{eqnarray*}}
\def\eeqnn{\end{eqnarray*}}
\def\eqn{\begin{eqnarray}}
\def\eeqn{\end{eqnarray}}

\def\bal{\begin{align}}
\def\eal{\end{align}}

\theoremstyle{plain}
\newtheorem{theorem}{Theorem}[section]

\numberwithin{equation}{section}

\def\prf{\begin{proof}}
\def\endprf{\end{proof}}

\begin{document}

\parskip=8pt

\title[Homogenous Fermi gas in a random medium]
{Boltzmann limit and quasifreeness for a homogenous Fermi gas in
a weakly disordered random medium}
\author[T. Chen]{Thomas Chen}
\address{T. Chen, Department of Mathematics,
Princeton University; and Department of Mathematics,
University of Texas at Austin.}
\email{tc@math.princeton.edu}

\author[I. Sasaki]{Itaru Sasaki}
\address{I. Sasaki, Department of Mathematics, Princeton University.}
\email{isasaki@math.princeton.edu}


\begin{abstract}
We discuss some basic aspects of the dynamics  
of a homogenous Fermi gas in a weak random potential, under 
negligence of the particle pair interactions. 
We derive the kinetic scaling limit for the momentum distribution function
with a translation invariant initial state and prove that it is
determined by a linear Boltzmann equation.
Moreover, we prove that if the initial state
is quasifree, then the time evolved state, averaged over the randomness,
has a quasifree kinetic limit.
We show that the momentum distributions determined by the
Gibbs states of a free fermion field are stationary solutions of 
the linear Boltzmann equation;
this includes the limit of zero temperature.
\end{abstract}

\maketitle

\section{Introduction}

We investigate the Boltzmann limit for the dynamics of
a quantized field of non-relativistic electrons in a disordered medium.
Our approach is based on 
techniques developed in 
\cite{Erd,ErdSalm,ErdSalmYau1,ErdSalmYau1-2,ErdYau} and \cite{Ch1,Ch2}, 
used to derive the Boltzmann limit for the one-particle Anderson model
at weak disorders;
see also \cite{Sp,LukSpo}.
We refer also to \cite{AiSiWa,Bou-1,Bou-2,Kl,RodSchl} for related works.

We consider a gas of fermions on the lattice $\LL:=[-\frac L2,\frac L2]^d\cap\Z^d$
in dimension $d\geq3$ and with periodic boundary conditions, for $L\gg1$.
We denote the dual lattice by $\LL^*=\LL/L$, and write
$\int dp \, \equiv \, \frac{1}{L^d}\sum_{p\in\LL^*}$
for brevity.
Letting $\Fo \, = \, \bigoplus_{n\geq0} \,\bigwedge_1^n \ell^2(\LL)$ denote the
Fock space accounting for scalar fermions on $\LL$, we denote the creation- and annihilation
operators by $a^+_p$, $a_p$, with $p \in\LL^*$,
satisfying the usual canonical anticommutation relations.

Let $\alg$ denote the $C^*$-algebra of bounded operators on $\Fo$.
We let $\omzl$ denote a translation invariant, normalized state on $\alg$,
which preserves the particle number (i.e., $\omzl(NA)=\omzl(AN)$ for all $A\in\alg$,
where $N=\sum_x a^+_x a_x$ is the number operator).

We consider the Hamiltonian
\eqn
	\Hom \, := \,
	\int dp \, E(p) \, a_p^+ \, a_p \, + \, \eta \, \Vom
\eeqn
which generates the dynamics of a free Fermi gas coupled to a random potential
\eqn
	\Vom \, := \, \sum_{x\in\LL} \omega_x \, a_x^+ \, a_x
\eeqn
where $\{\omega_x\}_{x\in\LL}$ are real Gaussian i.i.d. random variables, and
$0<\eta\ll1$ is a small coupling constant accounting for the disorder strength.
We assume that the kinetic energy function is given by
\eqn
	E(p) \, = \, \sum_{j=1}^d \cos(2\pi p_j) \,,
\eeqn
i.e., the Fourier multiplication operator determined by
the centered nearest neighbor Laplacian $(\Delta f)(x)=\sum_{|y-x|=1}f(y)$ on $\Z^d$.




We are interested in the long-time dynamics  
of the fermion field described by 
\eqn
	\omzlt( \, A \,) \, := \, \omzl( \, e^{it\Hom} \, A \, e^{-it\Hom} \, ) \,,
\eeqn 
where $A \in \alg$.
While we are neglecting the pair interactions between the electrons, 
the effective interaction between the particles through
their coupling to the random potential, and  
due to the Pauli principle remain significant.
We prove the following. In a time scale $t=\frac{T}{\eta^2}$ where $T>0$ denotes
a macroscopic time variable, we find, in the thermodynamic limit, that for all $T>0$
and for all test functions $f$, $g$ of  Schwartz class $\cS(\Tor^d)$,
\eqn
	\Omega_T^{(2)}(f;g) \, := \,
	\lim_{\eta\rightarrow0}\lim_{L\rightarrow\infty}
	\Exp[ \, \omzlT( \, a^+(f) \, a(g) \, ) \, ]
	\, = \, \int_{\Tor^d} dp \, F_T(p) \, \overline{f(p)} \, g(p) \,,
\eeqn
where $F_T(p)$ satisfies the linear Boltzmann equation
\eqn\label{eq-lBoltz-intro-1}
	\partial_T F_T(p) \, = \, 2 \, \pi \int du \, \delta( \, E(u)-E(p) \, )
	\, ( \, F_T(u) - F_T(p)\, )
\eeqn
with initial condition
$F_0(p)\,=\,\lim_{L\rightarrow\infty}\frac{1}{L^d}\omzl( \, a^+_p \, a_p \, ) $.
The proof is based on a generalization of methods due to Erd\"os and Yau
in \cite{ErdYau}, and extended in \cite{Ch1},
for the derivation of linear Boltzmann
equations from the random Schr\"odinger dynamics in the weakly disordered
1-particle Anderson model.

We observe that if $\omzl$ is the Gibbs distribution of the
free fermion field,
the corresponding momentum occupation density
(the Fermi-Dirac distribution)
\eqn\label{eq-Gibbs-mom-dist-1}
	F_0(p) \, = \, \frac{1}{1+e^{\beta (E(p) - \mu)}} \, ,
\eeqn
for inverse temperature $\beta$ 
and chemical potential $\mu$,
is a {\em stationary solution} of the linear Boltzmann equation
(\ref{eq-lBoltz-intro-1}), for all $\beta>0$. 
This is also valid in the zero temperature limit $\beta\rightarrow\infty$ where 
in the weak sense,
\eqn 
	\frac{1}{1+e^{\beta( E(p)-\mu)}} \, \rightarrow \, \chi[E(p)<\mu] \,,
\eeqn
which is nontrivial if $\mu>0$.
Erd\"os, Salmhofer and Yau have proved in their landmark work
\cite{ErdSalm,ErdSalmYau1,ErdSalmYau1-2}
that for a time $t$ beyond the kinetic scale $\eta^{-2}$, the effective dynamics
of a single electron is {\em diffusive}; i.e., in this time scale,
a wave packet evolves in position space
according to the solution of a heat equation.
Accordingly, we expect the Fermi-Dirac distribution to remain
a stationary solution in the diffusive limit, and the corresponding time scale
addressed in \cite{ErdSalm,ErdSalmYau1,ErdSalmYau1-2}.


We remark that the translation invariant model without the random
potential (i.e., $\eta=0$) but including the full repulsive particle pair interaction
is determined by the Hamiltonian
\eqn
	\widetilde H_\lambda \, := \, \int dp \, E(p) \, a_p^+ \, a_p
	\, + \,
	\lambda \sum_{x,y\in\LL} \, a_y^+ \, a_x^+ \, v(x-y) \, a_x \, a_y \,.
\eeqn
It is widely believed that in a time
scale $t=\frac{T}{\lambda^2}$, the momentum density
$F_T(p):=\lim_{\lambda\rightarrow0}
\lim_{L\rightarrow\infty}\frac{1}{L^d}\rho_{T/\lambda^2}( \, a_p^+a_p \, )$ 
for the dynamics generated by $\widetilde H_\lambda$ satisfies the
{\em Boltzmann-Uhlenbeck-Uehling equation}
\eqn\label{eq-intro-BUU-1}
	\partial_T F_T(p) & = & - 4\, \pi \int dp_1 \, dp_2 \, dq_1 \, dq_2 \,
	| \, \widehat v(p_1-q_1) - \widehat v(p_1-q_2) \, |^2 \,
	\delta( \, p-p_1 \, ) \,
	\nonumber\\
	&&\quad
	\delta( \, p_1 + p_2 - q_1 - q_2 \, )
	\, \delta( \, E(p_1) + E(p_2) - E(q_1) - E(q_2) \, ) 
	\nonumber\\
	&&\quad
	\Big[ \, F_T(p_1) F_T(p_2) \widetilde F_T(q_1) \widetilde F_T(q_2)
	\, - \,
	F_T(q_1) F_T(q_2) \widetilde F_T(p_1) \widetilde F_T(p_2) \, \Big] \,,
	\nonumber
\eeqn
where $\widetilde F_T(p):=1-F_T(p)$. The derivation of (\ref{eq-intro-BUU-1})
from the microscopic quantum dynamics
is an extremely challenging open problem; for some work in this direction,
see \cite{BCEP1,ErdSalmYau2,HoLan,Spo}.
We note that (\ref{eq-Gibbs-mom-dist-1}) is also an equilibrium solution
of (\ref{eq-intro-BUU-1}), which  
is ealily seen by noting that $\widetilde F_0(p)=e^{\beta(E(p)-\mu)} F_0(p)$.
This is a consequence of the fact that (\ref{eq-Gibbs-mom-dist-1}) is a
function of the kinetic energy $E(p)$ which is a {\em collision invariant} in both
(\ref{eq-lBoltz-intro-1}) and (\ref{eq-intro-BUU-1}).
As a matter of fact, any distribution of the form
$f(E(p))$ is stationary for (\ref{eq-lBoltz-intro-1}); on the other hand,
the special structure of (\ref{eq-Gibbs-mom-dist-1}) is necessary for it
to be a stationary solution of (\ref{eq-intro-BUU-1}).
For a combined Boltzmann limit of the coupled model with $\lambda,\eta>0$
(which is an  open problem) we believe that
the kinetic energy $E(p)$ will remain a collision invariant, and that
the momentum distribution (\ref{eq-Gibbs-mom-dist-1}) will remain a stationary solution
of the resulting Boltzmann equation.

A contextually related question is the problem of the stability of the Fermi sea for a
gas of interacting fermions. This is an important problem in mathematical
physics which has in recent years received much attention, especially due to
the landmark works of Feldman, Kn\"orrer, and Trubowitz 
summarized in \cite{FeldKnoTru}.

We also address the question how strongly the electrons are 
effectively correlated
through their interactions with the random potential. To this end, we assume
that $\omzl$ is number
preserving, homogenous, and {\em quasifree}.
That is, for any tuple of test functions $f_1,\dots,f_r$, $g_1,\dots,g_s$,
\eqn
	\omzl( \, a^+(f_1)\cdots a^+(f_r) \, a(g_1) \cdots a(g_s) \, )
	\, = \, \delta_{r,s} \, \det[ \, \omzl( \, a^+(f_j) \, a(g_\ell) \, )\, ]_{j,\ell=1}^r \,.
\eeqn
We consider the dynamics generated by $\Hom$, and observe that
since $\Hom$ is bilinear in $a^+,a$, the time evolved state
$\omzlt$ is almost surely quasifree.
However, the state averaged over the randomness is {\em not} quasifree,
\eqn
	\lim_{L\rightarrow\infty}\Exp[ \, \omzlt( \, f_1,\dots,f_r \, ; \, g_1,\dots,g_r \, ) \, ]
	\, \neq \,
	\det[ \, \lim_{L\rightarrow\infty} \Exp[ \, \omzlt( \, f_j \, ; \, g_\ell \, ) \, ]\, ]_{j,\ell=1}^r \,,
\eeqn
for any $t>0$ if $\eta>0$. This is not surprising because quasifreeness
is a nonlinear condition.
We prove that in the kinetic scaling limit stated above, the
limiting $2r$-correlation functions are quasifree,
\eqn
	\lefteqn{
	\Omega_T^{(2r)}( \, f_1,\dots,f_r \, ; \, g_1,\dots,g_r \, )
	}
	\nonumber\\
	&&\quad\quad\quad
	:=  \, \lim_{\eta\rightarrow0} \lim_{L\rightarrow\infty}
	\Exp[ \, \omzlT( \,  a^+(f_1)\cdots a^+(f_r) \, a(g_1) \cdots a(g_2) \, ) \, ]
	\nonumber\\
	&& \quad\quad\quad
	= \, \det[ \, \Omega_T^{(2)}( \, f_j \, ; \, g_\ell \, )\, ]_{j,\ell=1}^r
\eeqn
for any $r\in\N$.
The proof is based on an extension of the proof in \cite{Ch2} for the
1-particle Anderson model at weak disorders that the
random Schr\"odinger evolution converges in {\em arbitrary higher mean} 
to a linear Boltzmann evolution.
Quasifreeness of the $2r$-point correlation functions is a significant ingredient in
some approaches to the problem of quantum charge transport;
see for instance \cite{AsJaPi} and the references therein.


\section{Definition of the model}

We give a detailed definition of the mathematical model
described in the previous section. 
We consider a fermion gas in a finite box $\LL:=[-\frac L2,\frac L2]^d\cap\Z^d$ of side
length $L\gg1$, with periodic boundary conditions, in dimensions $d\geq3$.
We denote its dual lattice by $\Lambda_L^*:=\LL/L\subset\Tor^d$.
For the Fourier transform, we use the convention
\eqn	
	\widehat f(p) \, := \, \sum_{x\in\LL} \, e^{-2\pi i p\cdot x} \, f(x) \,,
\eeqn
where $p\in \Lambda_L^*$, and
\eqn
	f(x) \, = \, \frac{1}{L^d}\sum_{p\in \Lambda_L^*} \, e^{2\pi i p\cdot x} \, \widehat f(p)
\eeqn
for its inverse.
For brevity, we will use the notation
\eqn
	\int dp \, \equiv \, \frac{1}{L^d}\sum_{p\in \LL^*}
\eeqn
in the sequel, which recovers its usual meaning in the thermodynamic limit $L\rightarrow\infty$.

We denote the fermionic Fock space of scalar electrons by
\eqn
	\Fo \, = \, \bigoplus_{n\geq0} \, \Fo_n \,,
\eeqn
where
\eqn
	\Fo_0 \, = \, \C
	\; \; \; , \; \; \;
	\Fo_n \, = \, \bigwedge_1^n \, \ell^2(\LL) \; , \; n\geq1 \,.
\eeqn
We introduce creation- and annihilation operators $a^+_p$, $a_q$,
for $p$, $q\in\LL^*$,
satisfying the canonical anticommutation relations
\eqn
	a_p^+ \, a_q \, + \, a_q \, a_p^+ \, = \, \delta(p-q)
	\, := \,
	\left\{
	\begin{array}{lll}
	L^d & & {\rm if \; \; } p \, = \, q \\
	0 & & {\rm otherwise.}
	\end{array}
	\right.
\eeqn
%
We define the fermionic manybody Hamiltonian
\eqn\label{eq-H-def-1}
	\Hom \, := \, T \, + \, \eta \, \Vom \,
\eeqn
where
\eqn
	T \, = \, \int dp \, E(p) \, a_p^+ \, a_p
\eeqn
is the kinetic energy operator, and
\eqn
	\Vom \, := \, \sum_{x\in\LL} \omega_x \, a_x^+ a_x
\eeqn
couples the fermions to a static random potential;
$\{\omega_x\}_{x\in\LL}$ is a field of i.i.d. real-valued random variables which
we assume to be centered, normalized, and Gaussian for simplicity. Thus,
\eqn
	\Exp[ \, \omega_x \, ] \, = \, 0
	\; \; , \; \;
	\Exp[ \, \omega_x^2 \, ] \, = \, 1
\eeqn
for all $x\in\LL$.
Moreover, we assume that
\eqn
	E(p) \, = \, \sum_{j=1}^d \cos(2\pi p_j) \,,
\eeqn
which defines the Fourier multiplier corresponding to
the nearest neighbor Laplacian on $\Z^d$.

Let
\eqn
	N \, := \, \sum_{x\in\LL} a_x^+ a_x
\eeqn
denote the particle number operator.
It is clear that
\eqn
	[\Hom,N] \, = \, 0
\eeqn
holds.

Let $\alg$ denote the $C^*$-algebra of bounded operators on $\Fo$.
We consider the dynamics on $\alg$ given by
\eqn
	\alpha_t(A) \, = \, e^{it\Hom} \, A \, e^{-it\Hom}
\eeqn
generated by the random Hamiltonian $\Hom$.

\section{Statement of the main results}

We consider a normalized, translation-invariant, deterministic state
\eqn
	\omzl:\alg\longrightarrow\C \,.
\eeqn
We define the time-evolved state
\eqn
	\omzlt(A) \, := \, \omzl( \, e^{it\Hom} \, A \, e^{-it\Hom} \,) \,,
\eeqn
with $t\in\R$, and initial condition given by $\omzl$.
We particularly focus on the dynamics of the averaged two-point functions
\eqn
	\Exp[ \, \omzlt( \, a_p^+ a_q\, ) \, ] \,,
\eeqn
where $p,q\in\LL^*$.
Clearly,
\eqn
	\Exp[ \, \omzl( \, a_p^+ a_q\, ) \, ]
	\, = \,	\omzl( \, a_p^+ a_q\, )
	\, = \, \delta(p-q) \, \frac{1}{L^d} \, \omzl( \, a_p^+ a_p \, ) \,,
\eeqn
where
\eqn
	\delta(k) \, := \, L^d\delta_k \,,
\eeqn
and where
\eqn
	\delta_k \, = \,
	\left\{
	\begin{array}{ll}
	1 & {\rm if \; } p= q \;
	\\
	0 & {\rm otherwise}
	\end{array}
	\right.
\eeqn
denotes the Kronecker delta on the lattice $\LL^*$ (mod $\Tor^d$).
We remark that for fermions, 
\eqn\label{eq-density-bd-1} 
	0 \, \leq \, \frac{1}{L^d} \, \omzl( \, a_p^+ a_p \, )  \, \leq \, 1 \,,
\eeqn
since $\|a_p^{(+)}\|=L^{d/2}$ in operator norm, $\forall p\in\LL^*$.

\subsection{The Boltzmann limit}
We denote the microscopic time, position, and velocity variables by $(t,x,p)$,
and the corresponding macroscopic variables by $(T,X,V)=(\eta^2 t, \eta^2 x , v)$.
We prove that the momentum distribution $f_t(q)$ converges to a solution of a
linear Boltzmann equation in the limit $\eta\rightarrow0$.

\begin{theorem}\label{thm-main-1}
We assume that $\omzl$ is translation invariant.
Then, the averaged two-point functions are translation invariant,
\eqn\label{eq-thm-main-1-1}
	\Exp[\omzlt( \, a^+(f) a(g)\, )] \, = \,
	\int dp \, \overline{f(p)} \, g(p) \, \Exp[\omzlt( \, a_p^+ a_p \, )] \,,
\eeqn
(i.e., diagonal in $a_p^+,a_p$) for any $f,g\in\cS(\Tor^d)$ of Schwartz class,
and the thermodynamic limit
\eqn\label{eqn-omzlt-lim-1}
	\Omega_T^{(2;\eta)}(f;g) \, := \,
	\lim_{L\rightarrow\infty}  \Exp[\omzlT( \, a^+(f) \, a(g) \, )]
\eeqn
exists for all  $f,g\in\cS(\Tor^d)$, and $T>0$.

For any $T>0$ and   all  $f,g\in\cS(\Tor^d)$, the limit
\eqn
	\Omega_T^{(2)}(f;g) \, := \, \lim_{\eta\rightarrow0}\Omega_T^{(2;\eta)}(f;g)  \,
\eeqn
exists, and is the inner product of $f,g$ with respect to a Borel measure $F_T(p)dp$,
\eqn
	\Omega_T^{(2)}(f;g) \, = \, \int dp \, F_T(p) \, \overline{f(p)} \, g(p) \,,
\eeqn
where $F_T(V)$ satisfies the linear Boltzmann equation
\eqn\label{eq-lin-Boltz}
	\partial_T F_T(V) \, = \, 2\, \pi \, \int_{\Tor^d} dU \, \delta( \, E(U) - E(V) \, )
	\, ( \, F_T(U) - F_T(V)\, ) \,,
\eeqn
with initial condition
\eqn\label{eq-lin-Boltz-initc-1}
	F_0(p) \, = \, \lim_{L\rightarrow\infty}\frac{1}{L^d} \,
	\omzl( \, a_{p_{\LL^*}}^+ \, a_{p_{\LL^*}} \, )
\eeqn
%
for $p\in\Tor^d$, where $p_{\LL^*}:=Q_{\frac{1}{2L}}(p)\cap\LL^*$,
and $Q_\delta(p):=p+[-\delta,\delta)^d$.
\end{theorem}

We note that there exists a unique $p_{\LL^*}\in\LL^*$
such that $|p-p_{\LL^*}|\leq \frac 1{2L}$,  for every $p\in\Tor^d$.

An initial condition of particular interest is the Gibbs state
(with inverse temperature $\beta$ and chemical potential $\mu$) for a non-interacting
fermion gas,
\eqn\label{eq-free-Gibbs}
	\omzl(A) \, = \, \frac{1}{Z_{\beta,\mu}} \, \Tr ( \, e^{-\beta( T -\mu N)} A\, )
\eeqn
where
$Z_{\beta,\mu}:=\Tr( \, e^{-\beta( T -\mu N)} \, )$. The corresponding momentum
distribution function
\eqn
	\lim_{L\rightarrow\infty} \frac{1}{L^d} \, \omzl( \, a_p^+ a_p \, )
	\, = \, \frac{1}{1+e^{\beta( E(p)-\mu)}}
\eeqn
is a {\em fixed point} of the linear Boltzmann equation (\ref{eq-lin-Boltz}),
for all $\beta>0$, including the zero temperature limit $\beta\rightarrow\infty$ where 
in the weak sense,
\eqn 
	\frac{1}{1+e^{\beta( E(p)-\mu)}} \, \rightarrow \, \chi[E(p)<\mu] \,,
\eeqn
which is nontrivial if $\mu>0$.
We note that all our results in this paper remain valid in the limit $\beta\rightarrow\infty$.

\subsection{Quasifreeness}
We prove that if in addition to the conditions
formulated above, the initial state $\omzl$ is {\em quasifree},
then $\Exp[\rho_t]$, which is not quasifree for any $t>0$ if $\eta>0$, 
becomes quasifree in the kinetic scaling limit of Theorem \ref{thm-main-1}.

A state $\omzl$ is quasifree if for any normal ordered
product of creation- and annihilation operators
\eqn
	a_{p_1}^+ \cdots a_{p_r}^+ a_{q_1} \cdots a_{q_s} \,,
\eeqn
with arbitrary $r,s\in \N$ and $p_i,q_j\in\LL^*$,
\eqn
	\omzl( \, a_{p_1}^+ \cdots a_{p_r}^+ a_{q_1} \cdots a_{q_s} \, )
	\, = \, \delta_{r,s} \, \det\big[ \, \omzl( \, a^+_{p_i}a_{q_j} \, ) \, \big]_{1\leq i,j\leq r} \,.
\eeqn
That is, any higher order correlation function decomposes into the determinant
of the matrix of pair correlations.
In its most general form, a particle number conserving quasifree state
$\omzl:\alg\rightarrow\C$ can be written as
\eqn
	\omzl(A) \, := \, \frac{1}{Z_K} \, \Tr( \, e^{-K} A \, )
\eeqn
for $A\in\alg$, with
\eqn
	Z_K \, := \, \Tr( \, e^{-K} \, ) \,,
\eeqn
and
\eqn
	K \, = \, \int dp \, dq \, \kappa(p,q) \, a_p^+ \, a_q \,
\eeqn
bilinear in $a^+_p,a_q$; for a proof, see \cite{BaLiSo}.
We assume $K$ to be deterministic (with respect to $\{\omega_x\}_x$).

If in addition, translation invariance is imposed, such that
\eqn
	[K,T] \, = \, 0 \,
\eeqn
then
\eqn\label{eq-K-h-def-1}
	K \, = \, \int dp  \, \kker(p) \, a_p^+ \, a_p
\eeqn
is bilinear and diagonal in $a_p^+,a_p$.

Since $\Hom$ is bilinear in the creation- and annihilation operators, it
is immediately clear that
\eqn
	K(t) \, := \,  e^{it\Hom} \, K \, e^{-it\Hom}
\eeqn
is also bilinear in $a^+_p,a_q$. Therefore,
\eqn
	\omzlt(A) \, = \, \frac{1}{Z_K} \, \Tr( \, e^{-K(t)} A \, )
\eeqn
is quasifree with probability 1. However, since
quasifreeness is a {\em nonlinear} condition on determinants,
almost sure quasifreeness does {\em not}
imply that $\Exp[\omzlt(\, \cdot \, )]$ is quasifree.

In fact, $\Exp[\omzlt(\, \cdot \, )]$ is {\em not} quasifree for any $t>0$.

However, we prove in Theorem \ref{thm-main-2} below that
it possesses a kinetic scaling limit (in the sense of Theorem \ref{thm-main-1}) which is quasifree.

\begin{theorem}\label{thm-main-2}
Assume that $\rho_0$ is number conserving and quasifree,
and translation invariant.
Then, the following holds.
For any normal ordered monomial in creation- and annihilation operators,
\eqn
	a^+(f_1) \cdots a^+(f_r) \, a(g_1) \cdots a(g_r)  \,,
\eeqn
with $r,s\in \N$ and Schwartz class test functions $f_j,g_\ell\in\cS(\Tor^d)$,
and any $T>0$,
the macroscopic $2r$-point function
\eqn
	\lefteqn{
	\Omega_T^{(2r)}( \, f_1,\dots,f_r \, ; \, g_1,\dots, g_r \, )
	}
	\\
	&&\, := \, \lim_{\eta\rightarrow0} \lim_{L\rightarrow\infty}
	\Exp[\rho_{T/\eta^2}( \, a^+(f_1) \cdots a^+(f_r) \, a(g_1) \cdots a(g_r) \, )]
	\nonumber
\eeqn
exists and is quasifree,
\eqn
	\Omega_T^{(2r)}( \, f_1,\dots,f_r \, ; \, g_1,\dots, g_r \, )
	\, = \, \det\big[ \, \Omega_T^{(2)}( \,  f_i \, , \, g_j \, ) \, \big]_{1\leq i,j\leq r} \,.
\eeqn
The macroscopic 2-point function is the same as in Theorem \ref{thm-main-1},
\eqn
	\Omega_T^{(2)}( \, f \, ; \, g \, ) \, = \, \int dp  \, F_T(p) \, \overline{f(p)} \, g(p) \, ,
\eeqn
and $F_T(p)$ solves the
linear Boltzmann equation (\ref{eq-lin-Boltz})
with initial condition (\ref{eq-lin-Boltz-initc-1}).

\end{theorem}

We note that the assumption of translation invariance can easily be dropped.
However, we do not address inhomogenous Fermi gases in this text.


\section{Proof of Theorem {\ref{thm-main-1}}}

The proof of Theorem {\ref{thm-main-1}} is obtained from an extension of
the analysis in \cite{Ch1,ErdYau}.

\subsection{Duhamel expansion}

We consider the Heisenberg evolution of the creation- and annihilation
operators. We define
\eqn
	a_p(t) \, := \, e^{it\Hom} a_p e^{-it\Hom} \, ,
\eeqn
and
\eqn
	a(f,t) \, := \, e^{it\Hom} a(f) e^{-it\Hom} \,.
\eeqn
We make the key observation that
\eqn
	a(f,t) \, = \, a(f_t)
\eeqn
where $f_t$ is the solution of the 1-particle random Schr\"odinger equation
\eqn\label{eq-onepart-RSE-1}
	i\partial_t f_t \, = \,  \Hom^{(1)} \, f_t \, := \,  \Delta f_t \, + \, \eta \, \Vom^{(1)} f_t
\eeqn
with initial condition
\eqn\label{eq-onepart-RSE-2}
	f_0 \, = \, f \,.
\eeqn
Here, $\Delta$ denotes the nearest neighbor Laplacian on $\LL$, and $\Hom^{(1)}=\Hom|_{\Fo_1}$ is the
1-particle Anderson Hamiltonian at weak disorders studied in \cite{Ch1,Ch2,ErdYau}.
$\Vom^{(1)}=\Vom|_{\Fo_1}$ is the 1-particle
multiplication operator $(\Vom^{(1)}f)(x)=\omega_x f(x)$.

To prove (\ref{eq-onepart-RSE-1}), (\ref{eq-onepart-RSE-2}), 
we observe that since $\Hom$ is bilinear in $a^+,a$,
it follows that $a(f,t)$ is a linear superposition of annihilation operators.
Therefore, there exists a function $f_t$ such that $a(f,t)=a(f_t)$. In particular,
\eqn
	i\partial_t a(f_t) & = &
	[ \, \Hom \, , \, a(f_t) \, ]
	\nonumber\\
	& = & \int dp \, f_t(p) \, E(p) \, a_p
	\, + \, \eta \int dp \int du \, f_t(p) \, \homega(u-p) \, a_u
	\nonumber\\
	& = & a( \, \Delta f_t \, ) \, + \,  a( \, \eta \, \Vom^{(1)}f_t \, )\,,
\eeqn
and moreover, it is clear that $a(f,0)=a(f_0)=a(f)$. This implies 
(\ref{eq-onepart-RSE-1}), (\ref{eq-onepart-RSE-2}).

Thus,
\eqn
	\omzlt( \, a^+(f) \, a(g) \, ) & = & \omzl( \, a^+(f_t) \, a(g_t )\, )
	\nonumber\\
	& = & \int dp \, dq \, \omzl( \, a_p^+ \, a_q) \, \overline{f_t(p)} \, g_t(q)
	\nonumber\\
	& = & \int dp \, J(p) \, \overline{f_t(p)} \, g_t(p)
\eeqn
where
\eqn\label{eq-J-def-1}
	\omzl( \, a_p^+ \, a_q) \, = \, \delta(p-q) \, J(p)
\eeqn
due to translation invariance, with 
\eqn 
	0 \, \leq \, J(p) \, = \, \frac{1}{L^d}\omzl( \, a_p^+ a_p \, ) 
	\, = \, \frac{1}{1+e^{\kker(p)}}  \, \leq \, 1 \,,
\eeqn
cf. (\ref{eq-density-bd-1}); see (\ref{eq-K-h-def-1}) for the definition of $\kker(p)$.  
In particular, this implies (\ref{eq-thm-main-1-1}).

For $N\in\N$, which we determine later,
we expand $f_t$, $g_t$ into the truncated Duhamel series at level $N$,
\eqn
	f_t \, = \,   f_t^{(\leq N)} \, + \, f_t^{(>N)}  \,,
\eeqn
with
\eqn
	 f_t^{(\leq N)} \, := \, \sum_{n=0}^N f_t^{(n)}  \,,
\eeqn
and where the Duhamel term of $n$-th order (in powers of $\eta$) is given by
\eqn
	f_t^{(n)}(p)  & := & (i\eta)^n
	\int ds_0 \cdots ds_n \, \delta(t-\sum_{j=0}^n s_j)
	\\
	&&\quad\int dk_0\cdots dk_n \, \delta(p-k_0) \,
	\Big( \, \prod_{j=0}^n e^{is_j E(k_j)} \, \Big)
	\Big( \, \prod_{j=1}^n\homega(k_j-k_{j-1}) \, \Big) \, f(k_n)
	\nonumber\\
	&=& \eta^n \, e^{\e t} \, \int d\alpha \, e^{it\alpha}
	\, \int dk_0\cdots dk_n \, \delta(p-k_0)
	\\
	&&\quad
	\Big( \, \prod_{j=0}^n \frac{1}{ E(k_j) -\alpha-i\e} \, \Big)
	\Big( \, \prod_{j=1}^n\homega(k_j-k_{j-1}) \, \Big) \, f(k_n) \,.
	\nonumber
\eeqn
The remainder term is given by
\eqn
	f_t^{(>N)}  \, = \,  i\eta
	\int_0^t ds \, e^{i(t-s)\Hom} \,  \Vom^{(1)} \, f_t^{(N)}(s)  \,.
\eeqn
%
We choose
\eqn
	\e \, = \, \frac1t
\eeqn
so that the factor $e^{\e t}$ remains bounded for all $t$.
Accordingly,
\eqn\label{eq-Duh-exp-1}
	\omzlt( \, a^+(f)  \, a(g) \, ) \, = \,
	\omzl( \,  a^+(f_t) \, a(g_t)   \, )
	\, = \, \sum_{n,\tn=0}^{N+1} \omzlt^{(n,\tn)}(f;g)
\eeqn
where
\eqn
	\omzlt^{(n,\tn)}(f;g) \, := \,
	\omzl( \, a^{+}(f_t^{(n)}) \, a(g_t^{(\tn)}) \, )
\eeqn
if $n,\tn\leq N$,
and
\eqn
	\omzlt^{(n,N+1)}(f;g) & := &
	\omzl( \, a^{+}(f_t^{(n)}) \, a(g_t^{(>N)}) \, )  \, ,
	\nonumber\\
	\omzlt^{(N+1,\tn)}(f;g) & := &
	\omzl( \, a^{+}(f_t^{(>N)}) \, a(g_t^{(\tn)}) \, )
\eeqn
if $n\leq N$, respectively if $\tn\leq N$, and
\eqn
	\omzlt^{(N+1,N+1)}(f;g) \, := \,
	\omzl( \, a^{+}(f_t^{(>N)}) \, a(g_t^{(>N)}) \, ) \,.
\eeqn
In particular, for $n,\tn\leq N$,
\eqn\label{eq-rhontn-1}
	\lefteqn{
	\omzlt^{(n,\tn)}(f;g) \, = \,
	\eta^{n+\tn} \, e^{2 \e t} \, \int d\alpha \, d\talpha \, e^{it(\alpha-\talpha)}
	}
	\nonumber\\
	&&\int dk_0\cdots dk_n \int d\tk_0\cdots d\tk_{\tn}
	\, \overline{f(k_n)} \, g(\tk_{\tn}) \,
	\, J(k_0) \, \delta(k_0-\tk_{0}) \, )
	\nonumber\\
	&&
	\quad\quad
	\prod_{j=0}^n \frac{1}{ E(k_j) -\alpha-i\e}
	\, \prod_{\ell=0}^n \frac{1}{ E(\tk_\ell) -\talpha+i\e}
	\nonumber\\
	&&
	\quad\quad
	\prod_{j=1}^n\homega(k_j-k_{j-1})
	\, \prod_{\ell=1}^n\homega(\tk_{\ell-1}-\tk_{\ell})
	\,.
\eeqn
This expression, and the expressions involving $n$ and / or $\tn=N+1$,
are completely analogous to those appearing in the truncated Duhamel expansion
of the Wigner transform in \cite{Ch1,ErdYau}.

This permits us to use the methods of \cite{Ch1,ErdYau} to prove
Theorem {\ref{thm-main-1}}. We will here only sketch the strategy; for
the detailed proof, we refer to \cite{Ch1,ErdYau}. In our subsequent
discussion, we will compare the expressions appearing in the given
problem to those treated in \cite{Ch1,ErdYau}.

To begin with, we introduce a more convenient notation.
Clearly, if $n,\tn\leq N$, and $n+\tn$ is odd, $\Exp[\omzlt^{(n,\tn)}(p,q)]=0$.
Thus, we let
\eqn
	\bn \, := \, \frac{n  + \tn}{2} \, \in \, \N \,,
\eeqn
and we define $\{u_j\}_{j=0}^{2\bn+1}$ by
\eqn
	u_j \, := \,
	\left\{
	\begin{array}{cl}
	k_{n-j} & {\rm if \; } j\leq n \\
	\tk_{j-n-1} & {\rm if \; } j\geq n+1 \,.
	\end{array}
	\right.
\eeqn
Consequently,
\eqn\label{eq-rhontn-2}
	\lefteqn{
	\Exp[\omzlt^{(n,\tn)}(f;g)] \, = \,
	\eta^{2\bn} \, e^{2 \e t} \,
	\int d\alpha \, d\talpha \, e^{it(\alpha-\talpha)}
	}
	\\
	&&\int du_0\cdots du_{2\bn+1}
	\, \overline{f(u_0)} \, g(u_{2\bn+1})
	\, J(u_n) \, \delta(u_n-u_{n+1})
	\nonumber\\
	&&
	\quad\quad
	\prod_{j=0}^n \frac{1}{ E(u_j) -\alpha-i\e}
	\, \prod_{\ell=n+1}^{2\bn+1} \frac{1}{ E(u_\ell) -\talpha+i\e}
	\nonumber\\
	&&
	\quad\quad
	\Exp\Big[ \, \prod_{j=1}^{n}\homega(u_j-u_{j-1})
	\, \prod_{j=n+2}^{2\bn+1}\homega(u_j-u_{j-1}) \, \Big]
\eeqn
in these new variables, where we use that $\homega(u)^*=\homega(-u)$.

\subsection{Graph expansion}

Next, we take the expectation with respect to the random potential.
To this end, we introduce the set of {\em Feynman graphs} $\Gamma_{n,\tn}$,
with $n+\tn\in2\N$, as follows.

We consider two horizontal solid lines, which we refer to as
{\em particle lines}, joined by a distinguished vertex
which we refer to as the $\omzl$-vertex
(corresponding to the term $\omzl( \, a^+_{u_n} a_{u_{n+1}} \, )$.
On the line on its left, we introduce $n$ vertices, and on the line on its right, we insert $\tn$ vertices.
We refer to those vertices as {\em interaction vertices}, and enumerate them
from 1 to $2\bn$ starting from the left.
The edges between the interaction vertices are referred to as {\em propagator lines}.
We label them by the momentum variables $u_0$, ..., $u_{2\bn+1}$, increasingly indexed
starting from the left. To the $j$-th propagator line, we associate the resolvent
$\frac{1}{E(u_j)-\alpha-i\e}$ if $0\leq j \leq n$, and $\frac{1}{E(u_j)-\talpha+i\e}$ if $n+1\leq j \leq 2\bn+1$.
To the $\ell$-th interaction vertex (adjacent to the edges labeled by
$u_{\ell-1}$ and $u_{\ell}$), we associate the random potential $\homega(u_\ell-u_{\ell-1})$,
where $1\leq \ell\leq 2\bn+1$.

A {\em contraction graph} associated to the above pair of  particle lines
joined by the $\omzl$-vertex, and decorated by $n+\tn$ interaction vertices,
is the graph obtained by pairwise connecting interaction vertices by
dashed {\em contraction lines}.
We denote the set of all such contraction graphs
by $\Gamma_{n,\tn}$; it contains
\eqn
	|\Gamma_{n,\tn}| \, = \, (2\bn-1)(2\bn-3)\cdots 3\cdot 1
	\, = \, \frac{(2\bn)!}{\bn!2^{\bn}} \, = \, O(\bn!)
\eeqn
elements.

If in a given graph $\pi\in\Gamma_{n,\tn}$,
the $\ell$-th and the $\ell'$-th vertex are joined by a contraction line,
we write
\eqn
	\ell \, \sim_\pi \, \ell' \,,
\eeqn
and we associate the delta distribution
\eqn
	\delta(u_{\ell}-u_{\ell-1}-(u_{\ell'}-u_{\ell'-1}))
	\, = \,
	\Exp[ \, \homega(\, u_{\ell}-u_{\ell-1} \,) \, \homega(\, u_{\ell'}-u_{\ell'-1} \,) \, ]
\eeqn
to this contraction line.

\subsection{Classification of graphs}

For the proof of Theorem \ref{thm-main-1}, we classify Feynman graphs
as follows; see \cite{Ch1,ErdYau}, and Figure 1.
\begin{itemize}
\item
A subgraph consisting of one propagator line
adjacent to a pair of vertices $\ell$ and $\ell+1$, 
and a contraction line connecting them, i.e., $\ell\sim_\pi\ell+1$,  
where both $\ell$, $\ell+1$ are either $\leq n$
or $\geq n+1$, is called an {\em immediate recollision}.
\item
The graph $\pi\in\Gamma_{n,n}$ (i.e., $n=\tn=\bn$) with $\ell\sim_\pi 2n-\ell$ for all $\ell=1,\dots,n$,
is called a {\em basic ladder} diagram. The contraction lines are called {\em rungs} of the
ladder. 
We note that a rung contraction always has the form $\ell\sim_\pi\ell'$ with $\ell\leq n$ and $\ell'\geq n+1$.
Moreover, in a basic ladder diagram one always has that if
$\ell_1\sim_\pi\ell_1'$ and $\ell_2\sim_\pi\ell_2'$
with $\ell_1<\ell_2$, then $\ell_2'<\ell_1'$.
\item
A diagram $\pi\in\Gamma_{n,\tn}$ is called a {\em decorated ladder} if
any contraction is either an immediate recollision, or a rung contraction $\ell_j\sim_\pi\ell_j'$
with $\ell_j\leq n$ and $\ell_j'\geq n$ for $j=1,\dots,k$, and $\ell_1<\cdots<\ell_k$,
$\ell'_1>\cdots>\ell'_k$. 
Evidently, a basic ladder diagram is the special case of a decorated
ladder which contains no immediate recollisions (so that necessarily, $n=\tn$).
\item
A diagram $\pi\in\Gamma_{n,\tn}$ is called {\em crossing} if there is a pair of
contractions $\ell\sim_\pi\ell'$, $j\sim_\pi j'$, with $\ell<\ell'$ and
$j<j'$, such that $\ell<j$.
\item
A diagram $\pi\in\Gamma_{n,\tn}$ is called {\em nesting} if there is a subdiagram with
$\ell\sim_\pi\ell+2k$, with $k\geq1$, and either $\ell\geq n+1$ or $\ell+2k\leq n$,
with $j\sim_\pi j+1$ for $j=\ell+1,\ell+3,\dots,\ell+2k-1$.
The latter corresponds to a progression of $k-1$ immediate recollisions.
\end{itemize}
We note that any diagram that is not a decorated ladder contains at least a crossing
or a nesting subdiagram.
\\

\centerline{\epsffile{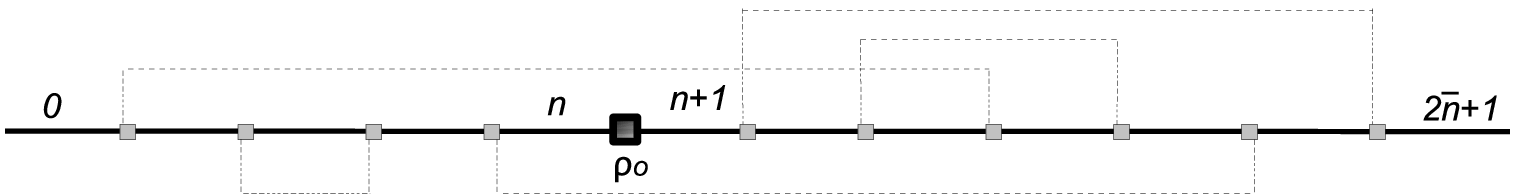} }

\noindent
Figure 1. An example of a Feynman graph, $\pi\in\Gamma_{n,\tn}$, with $n=4$, $\tn=6$.
The distinguished vertex is the $\omzl$-vertex.
\\

\subsection{Feynman amplitudes}

Next, we average (\ref{eq-rhontn-1}) with respect to the random potential. Accordingly,
$\Exp[ \, \prod\homega(u_\ell-u_{\ell-1}) \, ]$ splits into the sum of all possible
products of pair correlations, according to Wick's theorem (we recall that
$\{\omega_x\}$ are assumed to be i.i.d. Gaussian). This implies that
\eqn
	\Exp[ \, \omzlt^{(n,\tn)}(f;g) \, ] \, = \, \sum_{\pi\in\Gamma_{n,\tn}}
	\amp_\pi(f;g;\e;\eta)
\eeqn
with
\eqn\label{eq-rhontn-3}
	\lefteqn{
	\amp_\pi(f;g;\e;\eta)\, := \,
	\eta^{2\bn} \, e^{2 \e t} \,
	\int d\alpha \, d\talpha \, e^{it(\alpha-\talpha)}
	}
	\\
	&&\int du_0\cdots du_{2\bn+1}  \, \overline{f(u_0)} \, g(u_{2\bn+1})
	\, J(u_n) \, \delta(u_n-u_{n+1}) \,
	\nonumber\\
	&&
	\quad\quad
	\delta_\pi( \, \{u_j\}_{j=0}^{2\bn+1} \, ) \,
	\prod_{j=0}^n \frac{1}{ E(u_j) -\alpha-i\e}
	\, \prod_{\ell=n+2}^{2\bn} \frac{1}{ E(u_\ell) -\talpha+i\e}
	\,,
	\nonumber
\eeqn
and $\e=\frac1t$. Here,
\eqn
	\delta_\pi( \, \{u_j\}_{j=0}^{2\bn+1} \, )
	\, := \, \prod_{ \ell \sim_\pi \ell' }
	\delta( \, u_{\ell}-u_{\ell-1}-(u_{\ell'}-u_{\ell'-1}) \, )
\eeqn
is the product of the delta distributions associated to all contraction lines in $\pi$.
Moreover, we recall that
\eqn\label{eq-omzl-delta-1}
	\delta(\, u_n - u_{n+1} \, ) \, J(u_n)
	\, = \,
	\omzl( \, a^+_{u_n} a_{u_{n+1}} \, ) \,,
\eeqn
see (\ref{eq-J-def-1}).
We note that
\eqn\label{eq-mom-cons-1}
	u_0 \, - \, u_{2\bn+1} \, = \, 0 \,,
\eeqn
as one easily sees by summing up the arguments of all delta distributions.
This holds for any $n,\tn$
and again implies (\ref{eq-thm-main-1-1}).

We observe that the r\^ole of (\ref{eq-omzl-delta-1}) in (\ref{eq-rhontn-3})
is analogous to that of the rescaled Schwartz class function $J_{\e}$ in \cite{Ch1,ErdYau},
and that the test functions $f$, $g$ here correspond
to the initial state $\widehat\phi_0$ in \cite{Ch1,ErdYau}.

\subsection{Contribution from crossing and nesting diagrams}

The amplitude of any graph $\pi\in\Gamma_{n,\tn}$ that contains either a crossing or a nesting
can be estimated by
\eqn
	\lim_{L\rightarrow\infty} |\amp_\pi(f;g;\e;\eta)| \, \leq \, \| \, f \, \|_2
	\| \, g \, \|_2 \, \| \, J \, \|_\infty \,
	\e^{1/5} \, (\log\frac1\e)^4 (c\eta^2\e^{-1}\log\frac1\e)^{\bn} \,,
\eeqn
see \cite{Ch1,ErdYau}.
We note that similarly as in \cite{Ch1,ErdYau}, the bounds on all error terms
will only depend on the $L^2$-norm of the initial condition,
which in \cite{Ch1,ErdYau} is normalized by $\|\widehat\phi_0\|_2^2=1$.

The existence of the
the thermodynamic limit, as $L\rightarrow\infty$, is obtained
precisely in the same manner as in \cite{Ch1,Ch2}.
Let
\eqn
	\Gamma_{n,\tn}^{c-n} \, \subset \, \Gamma_{n,\tn}
\eeqn
denote the subset of diagrams of crossing or nesting type.
The number of graphs  in
\eqn
	\Gamma_{2\bn}^{c-n} \, := \, \bigcup_{n+\tn=2\bn} \, \Gamma_{n,\tn}^{c-n}
\eeqn
is bounded by $2^{\bn}\bn!$.

Thus, the sum of amplitudes associated to all crossing and nesting diagrams is bounded by
\eqn\label{eq-cnsum-bd-1}
	\lefteqn{
	\sum_{1\leq \bn \leq N} \sum_{\pi\in\Gamma_{2\bn}^{c-n}}
	\lim_{L\rightarrow\infty} |\amp_\pi(f;g;\e;\eta)|
	}
	\\
	&&\quad\quad\quad
	\, < \, (N+1)! \, \e^{1/5} \, (\log\frac1\e)^4 (c\eta^2\e^{-1}\log\frac1\e)^{N}
	\nonumber
\eeqn
noting that evidently, $\|f\|_2,\|g\|_2<C$ for $f,g $ of Schwartz class,
and recalling from (\ref{eq-density-bd-1}) that 
\eqn\label{eq-density-bd-2} 
	\| \, J \, \|_\infty \, \leq \, 1 \,,
\eeqn
which in particular is the case for
$J(p)=(1+e^{\beta(E(p)-\mu)})^{-1}$ associated to a Gibbs state of the
free Fermi field, for all $0\leq\beta\leq\infty$.

\subsection{Remainder term and time partitioning}

If at least one of the indices $n,\tn$ equals $N+1$, we first use
\eqn
	| \, \Exp[ \, \rho^{(N+1,\tn)}_t(f;g) \, ] \, |
	\, \leq \,  (\Exp[ \, \rho^{(\tn,\tn)}_t(g;g) \, ])^{1/2}
	\, (\Exp[ \, \rho^{(N+1,N+1)}_t(f;f) \, ])^{1/2}
\eeqn
by the Schwarz inequality (assuming without any loss of generality that $n=N+1$).
If $\tn\leq N$, the term $\Exp[ \, \rho^{(\tn,\tn)}_t(g;g) \, ]$
admits a bound of the form (\ref{eq-R1-bd-1}) below.

To bound $\Exp[ \, \rho^{(N+1,N+1)}_t(f;f) \, ]$, corresponding to
the remainder term in the Duhamel expansion,
we use the time partitioning method of \cite{ErdYau}; see also \cite{Ch1}.
To this end, we further expand the remainder term into $3N$ additional
Duhamel terms, and to subdivide the time integration interval $[0,t]$ into
$\kappa\in\N$ equal segments
\eqn
	[0,t] \, = \, \bigcup_{j=1}^\kappa[\tau_{j-1},\tau_j]
	\; \; , \; \;
	\tau_j=\frac{jt}{\kappa} \,,
\eeqn
whereby one obtains
\eqn
	f_t^{(>N)} \, = \, f_t^{(N,4N)}  \, + \, f_t^{(>4N)} \,,
\eeqn
where
\eqn\label{eq-ftN4N-def-1}
	f_t^{(N,4N)}  \,:= \, \sum_{j=1}^\kappa \, \sum_{n=N+1}^{4N-1} e^{i(t-\tau_j)\Hom^{(1)}}
	\widetilde f_{\tau_j}^{(n,N,\tau_{j-1})}  \, ,
\eeqn
with
\eqn
	\tf_s^{(n,N,\tau_{j-1})} \, := \,
	i \, \eta 
	\, \Vom^{(1)} f_s^{(n,N,\tau_{j-1})} \,,
\eeqn
and
\eqn
	\lefteqn{
	f_s^{(n,N,\tau)}(p) \, := \, (i\eta)^{n-N} \int_{\R^{n-N+1}}
	ds_0\cdots ds_{n-N} \, \delta(\sum_{j=0}^{n-N}s_j -(s-\tau))
	}
	\\
	&&\int du_0\cdots du_{n-N} \, \delta(p-u_0) \prod_{j=0}^{n-N} e^{is_j E(u_j)}
	\prod_{\ell=1}^{n-N}\homega(u_j-u_{j-1}) \, f(u_{n-N}) \,.
	\nonumber
\eeqn
Moreover,
\eqn
	f_t^{(>4N)} \, = \, \sum_{j=1}^\kappa e^{i(t-\tau_j)\Hom}
	\int_{\tau_{j-1}}^{\tau_j} ds \, e^{i(\tau_j-s)\Hom^{(1)}}
	\, \widetilde f_s^{(N,4N,\tau_{j-1})} \,.
	\nonumber
\eeqn
We note that writing (\ref{eq-ftN4N-def-1}) in the form 
$\sum_{j=1}^\kappa \, \sum_{n=N+1}^{4N-1}g_{n,j}$, we have 
\eqn
	\lefteqn{
	\omzlt( \, a^+(f_t^{(N,4N)}) \, a(f_t^{(N,4N)}) \, )
	}
	\nonumber\\
	& \leq & 
	\sum_{j,j'=1}^\kappa \sum_{n,n'=N+1}^{4N-1} \big| \, \omzl( \, a^+(g_{n',j'}) \, a(g_{n,j}) \, ) \, \big|
	\nonumber\\
	& \leq &\sum_{j,j'=1}^\kappa \sum_{n,n'=N+1}^{4N-1} 
	\frac12 \, \Big[ \, \omzl( \, a^+(g_{n',j'}) \, a(g_{n',j'}) \, ) 
	\, + \,  \omzl( \, a^+(g_{n,j}) \, a(g_{n,j}) \, )  \, \Big]
	\nonumber\\
	& \leq & \kappa^2 \, (3N)^2 \, \sup_{n,j}  \omzl( \, a^+(g_{n,j}) \, a(g_{n,j}) \, )  \,. 
\eeqn
Thus, by the Schwarz inequality,
\eqn
	\omzlt^{(N+1,N+1)}(f;f) \, \leq \, R_1(f, t) \, + \, R_2(f,t)
\eeqn
where
\eqn
 	R_1(f,t) \, := \, (3N)^2 \, \kappa^2 \,
	\sup_{\ontop{N< n < 4N}{1\leq j \leq\kappa}}
	\omzl( \, a^+(f_{\tau_j}^{(n,N,\tau_{j-1})}) \, a(f_{\tau_j}^{(n,N,\tau_{j-1})}) \, )
\eeqn
and
\eqn\label{eq-R2-def-1}
	R_2(f,t) \, := \, t^2 \, \sup_{1\leq j\leq \kappa}
	\sup_{s\in[\tau_{j-1},\tau_j]}
	\omzl( \, a^+(\tf_s^{ (N,4N,\tau_{j-1})} ) \,
	a(\tf_s^{(N,4N,\tau_{j-1})} )  \, ) \,.
\eeqn
By separating terms due to decorated ladders from those due to crossing and nesting diagrams,
one finds
\eqn\label{eq-R1-bd-1}
	\lefteqn{
	\lim_{L\rightarrow\infty}
	\Exp[ \, \omzl( \, a^+(f_{\tau_j}^{(n,N,\tau_{j-1})}) \, a(f_{\tau_j}^{(n,N,\tau_{j-1})}) \, )  \, ]
	}
	\nonumber\\
	&&\quad \quad \, = \,
	\Exp[ \, \int dp \, J(p) \, | \, f_{\tau_j}^{(n,N,\tau_{j-1})} (p) \, |^2 \, ]
	\nonumber\\
	&&
	\quad \quad
	\, \leq \, \| \, J \, \|_\infty \,
	\Exp[ \, \| \, f_{\tau_j}^{(n,N,\tau_{j-1})} \, \|_2^2 \, ]
	\\
	&&\quad \quad \, \leq \, \| \, f \, \|_2^2 \, \| \, J \, \|_\infty
	\, \Big[ \, \frac{ (c \, \e^{-1} \, \eta^2)}{(N!)^{1/2}}
	\, + \, \e^{1/5} \, (\log\frac1\e)^4 (c\eta^2\e^{-1}\log\frac1\e)^{8N} \, \Big] \,
	\nonumber
\eeqn
for $N<n<4N$ (see \cite{Ch1,ErdYau} for a detailed discussion).

For $n=4N$, the main issue is to control the large factor $t^2$ in (\ref{eq-R2-def-1}).
To this end, we observe that for a time integral on the interval $[\tau_{j-1},\tau_j]$ of
length $\frac{t}{\kappa}$, the parameter $\e=t^{-1}$ can
be replaced by $\kappa\e=(\frac{t}{ \kappa})^{-1}$. 
Therefore, one gets
\eqn\label{eq-rem-est-aux-2}
	\lefteqn{
	\lim_{L\rightarrow\infty}
	\Exp[ \, \omzl( \, a^+(\tf_s^{ (N,4N,\tau_{j-1})} ) \,
	a(\tf_s^{(N,4N,\tau_{j-1})} )  \, ) \, ]
	}
	\nonumber\\
	&&
	\quad \quad
	\, \leq \, \| \, J \, \|_\infty \,
	\Exp[ \, \| \, \tf_s^{ (N,4N,\tau_{j-1})} \, \|_2^2 \, ]
	\nonumber\\
	&&
	\quad \quad
	\, \leq \, \| \, f \, \|_2^2 \, \| \, J \, \|_\infty
	\, \Big[ \, \frac{((4N)!)}{\kappa^{2N}} \,
	(\log\frac1\e)^4 (c\eta^2\e^{-1}\log\frac1\e)^{8N} \, \Big] \, .
\eeqn
The gain of a factor $\kappa^{-2N}$ is crucial; it is sufficient to compensate for 
the factor $t^2$ in (\ref{eq-R2-def-1}), using the parameter choice given in
Section \ref{ssec-constants-1} below.

One obtains that if at least one of the indices $n$, $\tn$ equals $N+1$,
\eqn\label{eq-rem-est-aux-1}
	\lefteqn{
	\lim_{L\rightarrow\infty} |\Exp[\rho^{(n,\tn)}(f;f)]| \,
	\leq \, \| \, f \, \|_2^2 \, \| \, J \, \|_\infty
	\, \Big[ \,
	\frac{N^2 \, \kappa^2 \, (c \, \e^{-1} \, \lambda^2)}{(N!)^{1/2}}
	}
	\nonumber\\
	&&
	+ \, \Big( \, N^2 \, \kappa^2 \, \e^{1/5} \, + \, \e^{-2} \, \kappa^{-2N} \, \Big) \,
	((4N)!) \,
	(\log\frac1\e)^4 (c\lambda^2\e^{-1}\log\frac1\e)^{8N} \, \Big]\,,
\eeqn
where $\kappa$ remains to be chosen.
The first term on the right hand side of (\ref{eq-rem-est-aux-1})
bounds the contribution from all basic ladder diagrams contained in the
Duhamel expanded remainder term.
For a detailed discussion, we refer to \cite{Ch1,Ch2,ErdYau}.

\subsection{Choosing the constants}
\label{ssec-constants-1}
We recall from (\ref{eq-density-bd-2}) that $\| \, J \, \|_\infty\leq1$.
Moreover, $\| \, f \,\|_2$, $\| \, g \, \|_2<C$ for all test 
functions $f$, $g\in\cS(\Tor^d)$.
As in \cite{Ch1,Ch2,ErdYau}, we choose
\eqn\label{eq-param-choice-1}
	t \; = \; \frac1\e & = & \frac{T}{\eta^2}
	\nonumber\\
	N & = & \frac{\log\frac1\e}{10\log\log\frac1\e}
	\nonumber\\
	\kappa & = & (\log\frac1\e)^{15} \,.
\eeqn
Then,
\eqn
	\e^{-1/11} \; < \; N! & < & \e^{-1/10}
	\nonumber\\
	\kappa^N & > & \e^{-3/2}
\eeqn
and consequently,
\eqn\label{eq-rem-est-aux-1-1}
	(\ref{eq-cnsum-bd-1}) \, , \, (\ref{eq-rem-est-aux-2}) \, < \, \eta^{1/15}
\eeqn
and
\eqn\label{eq-rem-est-aux-1-2}
	(\ref{eq-rem-est-aux-1}) \, < \, \eta^{1/4}
\eeqn
for $\eta$ sufficiently small.
It follows that the sum of all crossing, nesting, and remainder terms
is bounded by $\eta^{1/20}$.

\subsection{Resummation of decorated ladder diagrams}

Let $\Gamma_{n,\tn}^{(lad)}\subset\Gamma_{n,\tn}$ denote the subset of all
decorated ladders based on $n+\tn$ vertices. Then, for $T>0$, let
\eqn
	\Omega_T^{(2;\eta)}(f;g) \, := \,
	\sum_{\bn=0}^{N(\e(T,\eta))} \sum_{n+\tn=2\bn} \sum_{\pi\subset \Gamma_{n,\tn}^{(lad)}}
	\lim_{L\rightarrow\infty} \amp_\pi(f;g;\e(T,\eta);\eta)
\eeqn
with $\e(T,\eta)=\frac{\eta^2}{T}$.
In the kinetic scaling limit $\eta\rightarrow0$ with
$t=\frac1\e=T/\eta^2$, one obtains
\eqn
	\Omega_T^{(2)}(f;g) \, := \,
	\lim_{\eta\rightarrow0}\Omega_T^{(2;\eta)}(f;g) \, = \,
	\int dp \, F_T(p) \, \overline{f(p)} \, g(p) \,,
\eeqn
where
\eqn\label{eq-FT-solexp-1}
	F_T(p) & := & \lim_{\eta\rightarrow0} F_T^{(\eta)}(p)
	\\
	& = & e^{-2\pi T \int du \, (E(u)-E(p))} \,
	\sum_{\bn=0}^\infty \int_{\R^{\bn+1}_+} dS_0\cdots dS_{\bn} \, \delta(T-\sum_{j=0}^{\bn} S_j)
	\nonumber\\
	&&
	\int du_0 \, \cdots \, du_n \, \delta(p-u_0)
	\Big( \, \prod_{j=1}^{\bn} 2 \, \pi \, \delta( \, E(u_j)-E(u_{j-1}) \, ) \, \Big) \, 
	F_0(u_n) \,,
	\nonumber
\eeqn
with initial condition
\eqn
	F_0(u) \, = \, \lim_{L\rightarrow\infty} J(u_{\LL^*})
	\, = \, \lim_{L\rightarrow\infty}\frac{1}{L^d} \, \omzl( \, a_{u_{\LL^*}}^+ a_{u_{\LL^*}} \, ) 
\eeqn
(for the definition of $u_{\LL^*}$, see Theorem \ref{thm-main-1}).
It can be straightforwardly verified that (\ref{eq-FT-solexp-1})
is a solution of the Cauchy problem
for the linear Boltzmann equation (\ref{eq-lin-Boltz}),
as asserted in Theorem \ref{thm-main-1}.

\section{Proof of Theorem {\ref{thm-main-2}}}

Because both $K$ (in the definition of $\omzl$) and the random Hamiltonian $\Hom$
are bilinear in $a^+,a$
(of the form $\int  du_1 \, du_2 \, k(u_1,u_2) \, a^+_{u_1} \, a_{u_2}$), the same
is true for
\eqn
	K(t) \, := \, e^{it\Hom} \, K \, e^{-it\Hom} \,,
\eeqn
with probability 1.
Therefore,
\eqn
	\omzlt( \, \cdot \, ) \, = \, \frac{1}{Z_K} \, \Tr( \, e^{-K(t)} \, ( \, \cdot \, )\, )
\eeqn
is quasifree with probability 1 (see, for instance, \cite{BaLiSo}).
Thus, for   $r,s\in\N$,
\eqn
	\omzlt( \, a^+(f_1) \cdots a^+(f_r) \, a(g_1) \cdots a(g_s) \, )
	\, = \, \delta_{r,s} \,
	\det\Big[ \,  \omzlt( \, a^+(f_j) a(g_\ell) \, ) \, \Big]_{j,\ell=1}^r \,,
\eeqn
where $f_j,g_\ell\in\cS(\Tor^d)$ belong to the Schwartz class.
In particular, we can set $r=s$.

We expand the determinant into
\eqn
	\lefteqn{
	\det\Big[ \,  \omzlt( \, a^+(f_j) \, a(g_\ell) \, ) \, \Big]_{j,\ell=1}^r
	}
	\\
	&&\, = \, \sum_{s\in S_r} (-1)^{\sign(s)} \prod_{j=1}^r
	\, \omzlt( \, a^+(f_j) \, a(g_{s(j)}) \, ) \,,
	\nonumber
\eeqn
where $S_r$ is the symmetric group of degree $r$. We claim that for $T>0$ and $t=\frac{T}{\eta^2}$,
and any choice of $f_j,g_\ell\in\cS(\Tor^d)$,
\eqn\label{eq-Lr-convbd-1}
	\lim_{L\rightarrow\infty}\Big| \, \Exp \Big[ \, \prod_{j=1}^r
	\, \omzlT( \, a^+(f_j) a(g_{s(j)}) \, ) \, \Big]
	\, - \,
	\prod_{j=1}^r \, \Exp[ \, \omzlT( \, a^+(f_j) \, a(g_{s(j)}) \, ]\, \Big|
	\, < \, \eta^{\delta} \,,
\eeqn
for a constant $\delta>0$ independent of $r$, $s\in S_r$, $\eta$, and $T$,
and for $\eta>0$ sufficiently small.
This immediately implies that, for every fixed $r<\infty$,
\eqn
	\lefteqn{
	\lim_{L\rightarrow\infty}
	\Big| \, \Exp\Big[\, \omzlT( \, a^+(f_1) \cdots a^+(f_r) \, a(g_1) \cdots a(g_r) \, )
	\, \Big]
	}
	\\
	&&\quad \quad \quad \, - \,
	\det\Big[ \,  \Exp[ \, \omzlT( \, a^+(f_j) a(g_\ell) \, ) \, ] \, \Big]_{j,\ell=1}^r
	\, \Big| \, < \, r! \, \eta^{\delta}
	\nonumber
\eeqn
converges to zero as $\eta\rightarrow0$.

This implies that
\eqn
	\lefteqn{
	\Omega_T^{(2r)}( \, f_1,\dots,f_r \, ; \, g_1,\dots, g_r  \, )
	}
	\\
	&& \, := \, \lim_{\eta\rightarrow0} \lim_{L\rightarrow\infty}
	\Exp[\rho_{T/\eta^2}( \, a^+(f_1) \cdots a^+(f_r) \, a(g_1) \cdots a(g_r) \, ) ]
	\nonumber
\eeqn
is quasifree, i.e.,
\eqn
	\Omega_T^{(2r)}( \, f_1,\dots,f_r \, ; \, g_1,\dots, g_r  \, )
	\, = \,  \det\big[ \, \Omega_T^{(2)}( \, f_i \, ; \, g_j \, ) \, \big]_{1\leq i,j\leq r} \,,
\eeqn
where
\eqn
	\Omega_T^{(2)}( \, f \, ; \, g \, ) \, = \, \int dp \, F_T(p) \, \overline{f(p)} \, g(p) \,.
\eeqn
The function $F_T(p)$ solves the
linear Boltzmann equation with initial condition $F_0(p)$,
as given in Theorem \ref{thm-main-1}.


\subsection{Proof of (\ref{eq-Lr-convbd-1})}

The inequality (\ref{eq-Lr-convbd-1}) follows from a straightforward application of
the main results in \cite{Ch2} where we refer for details.
In this section, we shall only outline the strategy.
The expectation
\eqn\label{eq-prod-adagga-r-1}
	\lim_{L\rightarrow\infty} \, \Exp \Big[ \, \prod_{j=1}^r
	\, \omzlt( \, a^+(f_j) \, a(g_{s(j)}) \, ) \, \Big]
\eeqn
can be represented by a graph expansion as follows.
We expand each of the factors
\eqn
	\omzlt( \, a^+(f_j) a(g_{s(j)}) \, )
	\, = \,
	\sum_{n,\tn=1}^{N+1} \int dp \, J(p) \, \overline{f_{j,t}^{(n)}(p)} \, g_{s(j),t}^{(\tn)}(p)
\eeqn
separately into a truncated Duhamel series of level $N$,
using the same definitions as in (\ref{eq-Duh-exp-1}).
For the remainder term (where at least one of the indices $n,\tn$ equals $N+1$),
we subdivide the time integration interval $[0,t]$
into $\kappa$ pieces of length $\frac{t}{\kappa}$.

For the expectation (\ref{eq-prod-adagga-r-1}), we introduce the
following extension of the classes of Feynman graphs discussed
for the proof of Theorem \ref{thm-main-1}, see also Figure 2.
For $r>1$, we consider $r$ particle lines parallel to one another,
each containing a distinguished $\omzl$-vertex separating it into
a left and a right part. Enumerating them from 1 to $r$, the
$j$-th particle line contains $n_j$ interaction vertices on
the left of the $\omzl$-vertex, and $\tn_j$ interaction vertices
on its right. We note that for $r>1$, only $\sum_{j=1}^r (n_j+\tn_j)$
has to be an even number, but not each individual
\eqn
	\hn_j \, := \, n_j \, + \, \tn_j \,.
\eeqn
On the $j$-th interaction line, we label the propagator lines
by momentum variables $u_0^{(j)},\dots,u_{ \hn_j+1}^{(j)}$,
with indices increasing from the left.
\\

\centerline{\epsffile{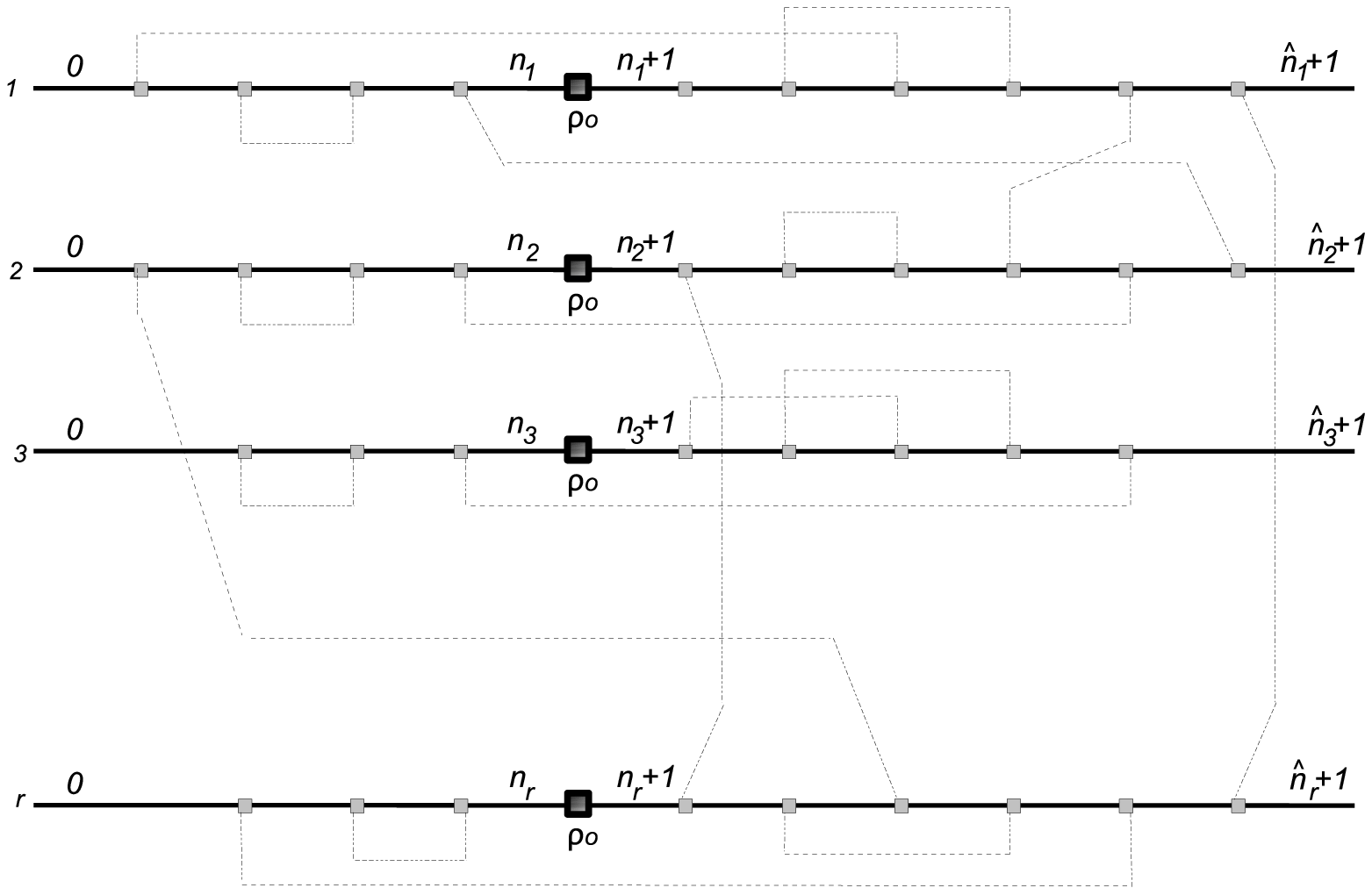} }

\noindent
Figure 2. Order $r$ Feynman graph. The particle line indexed by $j=3$ is 
disconnected.
\\

A {\em contraction graph} of degree $\{(n_j,\tn_j)\}_{j=1}^r$
is obtained by connecting pairs of interaction vertices by
contraction lines. We denote the set of contraction graphs of degree
$\{(n_j,\tn_j)\}_{j=1}^r$ by $\Gamma_{\{(n_j,\tn_j)\}_{j=1}^r }$.
If the $\ell$-th vertex on the $j$-th particle line is connected
by a contraction line to the $\ell'$-th vertex on the $j'$-th
particle line, we write
\eqn
	(j;\ell)  \, \sim_{\pi} \, (j';\ell') \,.
\eeqn
To a graph $\pi\in\Gamma_{\{(n_j,\tn_j)\}_{j=1}^r }$, we associate
the {\em Feynman amplitude}
\eqn\label{eq-rhontn-4}
	\lefteqn{
	\amp_\pi(\{f_j,g_{s(j)}\};\eta;T)\, := \,
	\eta^{2\sum_{1\leq j \leq r}( n_j+\tn_j)} \, e^{2 r \e t} \,
	\prod_{j=1}^r\int d\alpha_j \, d\talpha_j \, e^{it(\alpha_j-\talpha_j)}
	}
	\nonumber\\
	&&\int du_0^{(j)}\cdots du_{\hn_j+1}^{(j)}
	\, \overline{f_j(u_0^{(j)})} \, g_{s(j)}(u_{\hn_j+1}^{(j)})
	\, J(u_{n_j}^{(j)}) \, \delta( \, u_{n_j}^{(j)} - u_{n_j+1}^{(j)} \, ) \,
	\\
	&&
	\quad\quad
	\delta_\pi( \, \{u_i^{(j)}\}_{i=0}^{\hn_j+1} \, ) \,
	\prod_{\ell=0}^{n_j} \frac{1}{ E(u_\ell^{(j)}) -\alpha_j-i\e}
	\, \prod_{\ell'=n_j+2}^{\hn_j} \frac{1}{ E(u_{\ell'}^{(j)}) -\talpha_j+i\e}
	\,,
	\nonumber
\eeqn
where
\eqn
	\e \, = \, \frac 1t \, = \frac{\eta^2}{T}
\eeqn
for $T>0$.
The delta distribution
\eqn
	\delta_\pi( \, \{u_j^{(j)}\}_{j=0}^{\hn_j+1} \, )
	\, = \, \prod_{(j;\ell)  \, \sim_{\pi} \, (j';\ell')}
	\delta( \, u_{\ell}^{(j)}-u_{\ell-1}^{(j)} - ( u_{\ell'}^{(j')}-u_{\ell'-1}^{(j')} ) \, )
\eeqn
is the product of delta distributions associated to all contraction
lines in $\pi$.
\\

\subsubsection{Completely disconnected graphs}
The subclass
\eqn
	\Gamma_{\{(n_j,\tn_j)\}_{j=1}^r }^{disc}
	\, \subset \, \Gamma_{\{(n_j,\tn_j)\}_{j=1}^r }
\eeqn
of {\em completely disconnected} graphs of degree $\{(n_j,\tn_j)\}_{j=1}^r$
consists of those graphs in which
contraction lines only connect interaction vertices on the
same particle line.

It is clear that
\eqn
	\lefteqn{
	\lim_{L\rightarrow\infty}
	\sum_{\ontop{0\leq n_j,\tn_j\leq N}{ j=1,\dots,r } }
	\sum_{\pi\in \Gamma_{\{(n_j,\tn_j)\}_{j=1}^r }^{disc}}
	\amp_\pi(\{f_j,g_{s(j)}\};\eta;T)
	}
	\\
	&& \, = \,
	\lim_{L\rightarrow\infty}
	\prod_{j=1}^r \sum_{n_j,\tn_j=1}^N \Exp[ \, \int dp \, J(p) \,
	\overline{f_{j,T/\eta^2}(p)} \, g_{s(j),T/\eta^2}(p) \, ]
	\nonumber\\
	&& \, = \,
	\lim_{L\rightarrow\infty}
	\prod_{j=1}^r \big( \, \Exp[ \, \omzlT( \, a^+(f_j) a(g_{s(j)}) \, ) \, ] \,
	\, + \, O(\eta^\delta) \, \big) \,,
\eeqn
according to our proof of Theorem \ref{thm-main-1}. The term of order $O(\eta^\delta)$
accounts for the remainder term associated to the $j$-th particle line
(i.e., the terms involving $\Exp[ \, \omzlT^{(n_j,\tn_j)}(p,q) \, ] $
where at least one of the indices $n_j,\tn_j$ equals $N$).
Thus, for any fixed $r\in\N$, we obtain
\eqn
	&&\lim_{\eta\rightarrow0}\lim_{L\rightarrow\infty}
	\sum_{\ontop{0\leq n_j,\tn_j\leq N }{ j=1,\dots,r} }
	\sum_{\pi\in \Gamma_{\{(n_j,\tn_j)\}_{j=1}^r }^{disc}}
	\amp_\pi(\{f_j,g_{s(j)}\};\eta;T)
	\nonumber\\
	&&\quad \quad \quad \, = \,
	\prod_{j=1}^r \Omega_T^{(2)}( \, f_j \, ; \, g_{s(j)} \, ) \,.
\eeqn
That is, the sum over completely disconnected graphs yields the corresponding product of averaged
2-point functions in the kinetic scaling limit.

\subsubsection{Non-disconnected graphs}

We refer to the complement of the set of completely disconnected graphs in $\Gamma_{\{(n_j,\tn_j)\}_{j=1}^r }$,
\eqn
	\Gamma_{\{(n_j,\tn_j)\}_{j=1}^r }^{n-d} \, := \,
	\Gamma_{\{(n_j,\tn_j)\}_{j=1}^r } \, \setminus \, \Gamma_{\{(n_j,\tn_j)\}_{j=1}^r }^{disc} \,,
\eeqn
as the set of {\em non-disconnected graphs}.
It remains to prove that the sum over non-disconnected graphs,
combined with the remainder terms, can be bounded by $O(\eta^\delta)$, for $L$ sufficiently
large.

The condition required in \cite{Ch2} for the estimate analogous
to (\ref{eq-Lr-convbd-1}) to hold is that for the initial condition $\phi_0$
(corresponding to the test functions $f_j$, $g_\ell$ in our case) of the
random Schr\"odinger evolution studied in \cite{Ch2},
a "concentration of singularity condition" is satisfied
(that is, singularities in momentum space are not too much
``spread out'' in the limit $\eta\rightarrow0$). It states that in frequency space $\Tor^d$,
\eqn
	\widehat\phi_0 \, = \, \widehat\phi_0^{(reg)} \, + \, \widehat\phi_0^{(sing)} \,,
\eeqn
where
\eqn
	\| \,  \widehat\phi_0^{(reg)} \, \|_\infty \, < \, c
\eeqn
and
\eqn
	\| \, |\widehat\phi_0^{(sing)}| * |\widehat\phi_0^{(sing)}| \, \|_2  \, < \, c' \, \eta^{3/2}
\eeqn
are satisfied uniformly in $L$, as $L\rightarrow\infty$.

In the present case, we have to require that
$f_j$, $g_\ell$ satisfy the concentration of singularity condition.
This is, however, evidently fulfilled since
$f_j$, $g_\ell$ are $\eta$-independent Schwartz class functions
(in contrast, the initial states considered in \cite{Ch2} are of WKB type, and
scale non-trivially with $\eta$.)

It is proven in \cite{Ch2} that the amplitude of every non-disconnected graph
with $n_j,\tn_j\leq N$ for $j=1,\dots,r$, is bounded by
\eqn
	\lefteqn{
	\sup_{ \pi \in \Gamma_{\{(n_j,\tn_j)\}_{j=1}^r }^{n-d} }
	\big| \, \amp_\pi(\{f_j,g_{s(j)}\};\eta;T) \, \big|
	}
	\\
	&& \quad \quad	\, < \,
	\e^{1/5} \, (c \, \eta^2 \, \e^{-1} \, \log\frac1\e)^{\frac{r}{2} \sum_{j=1}^r\hn_j}
	(\log\frac1\e)^{4r} \,,
\eeqn
where we recall that $\e \, = \, \frac 1t \, = \frac{\eta^2}{T}$ for $T>0$.
This key estimate is a factor $\e^{1/5}$ smaller than the bound on
the sum of disconnected graphs; this improvement is obtained from exploiting
the existence of at least one contraction line that connects two different
particle lines; see \cite{Ch2}.

The number of non-disconnected graphs is bounded by
\eqn
	\big| \, \Gamma_{\{(n_j,\tn_j)\}_{j=1}^r }^{n-d} \, \big|
	\, \leq \, (\sum_{j=1}^r \hn_j) !  \, \leq \, (2rN)!
\eeqn
where $\hn_j=n_j+\tn_j$. Therefore, the sum of amplitudes of all non-disconnected graphs with
$0\leq n_j,\tn_j\leq N$ is bounded by
\eqn\label{eq-ndampl-bd-1}
	\lefteqn{
	\sum_{1\leq j\leq r}
	\, \, \sum_{0 \leq n_j,\tn_j \leq N } \,
	\sum_{\pi \in \Gamma_{\{(n_j,\tn_j)\}_{j=1}^r }^{n-d} }
	\big| \, \amp_\pi(\{f_j,g_{s(j)}\};\eta;T) \, \big|
	}
	\\
	&& \quad \quad
	\leq \,  ((2rN) !)^2
	\e^{1/5} \, (c \, \eta^2 \, \e^{-1} \, \log\frac1\e)^{rN}
	(\log\frac1\e)^{4r} \,.
\eeqn
Here we have estimated the sum over pairs $0 \leq n_j,\tn_j \leq N$, $1\leq j\leq r$,
by another factor $(2rN) !$, 
since $\#\{(n_j,\tn_j)\}_{j=1}^r \, | \, \sum_j\hn_j = m \, \} \leq m!$.

\subsubsection{Duhamel remainder term}

In case at least one of the indices $n_j$ or $\tn_j$ equals $N+1$,
the following argument can be applied. Clearly, from a H\"older estimate of
the form
$\|h_1\cdots h_s\|_1\leq\|h_1\|_s\cdots\|h_s\|_s$ with respect to $\Exp$,
we have
\eqn\label{eq-rem-Lr-est-1}
	\Big| \, \Exp\Big[ \, \prod_{j=1}^r \, 
	\omzlt^{(n_j,\tn_j)}(f;g)	
	\, \Big] \, \Big|
	\, \leq \,  \, \prod_{j=1}^r \, \Exp[ \, | \,
	\omzlt^{(n_j,\tn_j)}(f;g)
	\, |^{2r} \, ]^{\frac{1}{2r}} \,.
\eeqn	
Here, we have used an exponent $2r$ instead of $r$ because then,
even for $r$ odd, an
absolute value of the form $|z|^{2r}$ can be replaced by a product of the form
$\overline{z}^r z^r$, where $z\in\C$.

We make a choice of constants
\eqn\label{eq-param-choice-2}
	t \; = \; \frac1\e & = & \frac{T}{\eta^2}
	\nonumber\\
	N & = & \frac{\log\frac1\e}{10 r \log\log\frac1\e}
	\nonumber\\
	\kappa & = & (\log\frac1\e)^{15 r} \,,
\eeqn
similarly as in Section \ref{ssec-constants-1}
of the proof of Theorem \ref{thm-main-1}.

If $n_j$ or $\tn_j$ equals $N+1$, we can use the bounds (\ref{eq-rem-est-aux-1-1})
and (\ref{eq-rem-est-aux-1-2}).

If both $n$, $\tn \leq N$, we use the a priori bound
\eqn\label{eq-aprioribd-1}
	\lefteqn{
	\sum_{n+\tn=2\bn} \sum_{\pi\in\Gamma_{n,\tn} }
	\lim_{L\rightarrow\infty}\Exp[ \, | \,
	\omzlt^{(n ,\tn )}(f;g)
	\, |^{2r} \, ]^{\frac{1}{2r}}
	}
	\\
	&&\quad\quad\quad
	\, < \, \Big[ \, \sum_{\ell=0}^{2r} \Big(\ontop{2r}{\ell}\Big)
	\Big( \frac{(c\eta^2\e^{-1})^{\bn}}{(\bn!)^{1/2}}\Big)^{\ell} \, \times
	\nonumber\\
	&&
	\quad\quad\quad\quad\quad\quad
	\times \, \e^{1/5} \,  ((2r-\ell)\bn)! \,
	\Big( \, (\log\frac1\e)^4 (c\eta^2\e^{-1}\log\frac1\e)^{ \bn} \, \Big)^{2r-\ell}
	\Big]^{\frac{1}{2r}}
	\nonumber\\
	&&\quad\quad\quad
	\, < \,  \frac{(c\eta^2\e^{-1})^{\bn}}{(\bn!)^{1/2}}
	\, + \,  \eta^{\frac{1}{10}} \,
\eeqn
The factor $\frac{(c\eta^2\e^{-1})^{\ell\bn}}{(\bn!)^{\ell}}$ in $[\cdots]$ accounts for $\ell$
basic ladders on $\ell$ copies of $\Gamma_{n,\tn}^{disc}$, while the
remaining factor  accounts for all other (not necessarily non-disconnected) contractions on the remaining
$2r-\ell$ particle lines;
for details, see \cite{Ch1,Ch2,ErdYau}.

Let us without any loss of generality assume that $n_1=N+1$. Then, keeping
$n_1$ fixed and summing over the remaining indices $\tn_1$ and $n_j$, $\tn_j$, with $j=2,\dots,r$,
we find
\eqn
	\lefteqn{
	\sum_{\ontop{0\leq n_2, n_j,\tn_j\leq N+1}{j=2,\dots,r}}
	\prod_{j=1}^r \, \Exp[ \, | \,
	\omzlt^{(n_j,\tn_j)}(f;g)
	\, |^{2r} \, ]^{\frac{1}{2r}}
	}
	\nonumber\\
	&& \quad\quad\quad
	< \, \eta^{\frac{1}{15}} \, \Big[ \, \sum_{\bn=0}^N \frac{(c\eta^2\e^{-1})^{\bn}}{(\bn!)^{1/2}}
	\, + \,  \eta^{\frac{1}{10}} \, \Big]^{2r-1}
\eeqn
where the factor $\eta^{\frac{1}{15}}$ accounts for the remainder term indexed
by $n_1=N+1$.
We conclude that the sum over all terms (\ref{eq-rem-Lr-est-1}) which
contain at least one $n_j$ or $\tn_j$ equalling $N+1$ (i.e., which
contain at least one Duhamel remainder term) can be bounded by
\eqn
	C^r \, \eta^{\frac{1}{15}}
\eeqn
for a constant $C$ independent of $\eta$ and $r$.

Combined with
\eqn
	(\ref{eq-ndampl-bd-1}) \, < \, \eta^{\frac{1}{20}} \,,
\eeqn
which one easily verifies, this completes the proof of
Theorem \ref{thm-main-2}.
For more details addressing the arguments outlined here,
we refer to \cite{Ch2}.

\subsection*{Acknowledgements}
The authors are much indebted to H. Spohn.
We are grateful to one of the referees for detailed and very helpful comments.
T.C. thanks I. Rodnianski for very helpful discussions.
The work of T.C. was supported by NSF grants DMS-0524909 and DMS-0704031.
I.S. was supported by JSPS fellowship grant 18-4218.

\parindent=0pt


\begin{thebibliography}{99}
\bibitem{AiSiWa}
M. Aizenman, R. Sims, S. Warzel,
{\em Absolutely continuous spectra of quantum tree graphs with weak disorder}.  
Comm. Math. Phys. {\bf 264} (2) 371--389  (2006).
\bibitem{AsJaPi}
W. Aschbacher, V. Jaksic, Y. Pautrat, C.-A. Pillet,
{\em Transport properties of quasi-free fermions},  J. Math. Phys. {\bf 48}, no. 3   (2007)
\bibitem{BaLiSo}
V. Bach, E.H. Lieb, J.P. Solovej,
{\em Generalized Hartree-Fock theory and the Hubbard model},
J. Statist. Phys. {\bf 76} (1-2), 3--89 (1994).
\bibitem{BCEP1}
D. Benedetto, F. Castella, R. Esposito, M. Pulvirenti, {\em Some considerations on the
derivation of the nonlinear quantum Boltzmann equation}, J. Stat. Phys. {\bf 116},  no. 1-4, 381--410 (2004).
\bibitem{Bou-1} 
J. Bourgain, 
{\em On random Schr\"odinger operators on $\Z^2$}, 
Discrete Contin. Dyn. Syst. {\bf 8}, no. 1, 1-15, (2002).
\bibitem{Bou-2} J. Bourgain,  {\em Random lattice Schr\"odinger operators with
decaying potential: Some higher dimensional phenomena}, Springer LNM, Vol
1807 (2003), 70-98.
\bibitem{Ch1}
T. Chen, {\em Localization lengths and Boltzmann limit for the Anderson model at small disorders in dimension 3},
J. Stat. Phys., {\bf 120} (1-2), 279-337 (2005).
\bibitem{Ch2}
T. Chen, {\em Convergence in higher mean of a random Schr\"odinger to a linear Boltzmann evolution},
Comm. Math. Phys., {\bf 267}, 355-392 (2006).
\bibitem{Erd}
L. Erd\"os, {\em Linear Boltzmann equation as the scaling limit of
the Schr\"odinger evolution coupled to a phonon bath}, J. Stat. Phys.
{\bf 107} (5), 1043-1127 (2002).
\bibitem{ErdSalm}
L. Erd\"os, M. Salmhofer, {\em Decay of the Fourier transform of surfaces with vanishing curvature},
Math. Z. {\bf 257} (2), 261--294  (2007).
\bibitem{ErdSalmYau1}
L. Erd\"os, M. Salmhofer, H.-T. Yau, {\em Quantum diffusion for the Anderson model in the scaling limit},
Ann. Henri Poincar\'e, {\bf 8} (4), 621--685  (2007).
\bibitem{ErdSalmYau1-2}
L. Erd\"os, M. Salmhofer, H.-T. Yau,
{\em Quantum diffusion of the random Schr\"odinger evolution in the scaling limit. II. The recollision diagrams.}
Comm. Math. Phys. {\bf 271} (1), 1--53  (2007).
\bibitem{ErdSalmYau2}
L. Erd\"os, M. Salmhofer, H.-T. Yau, {\em On the quantum Boltzmann equation}, J. Stat. Phys., {\bf 116} (114),
367--380 (2004).
\bibitem{ErdYau}
L. Erd\"os, H.-T. Yau, {\em Linear Boltzmann equation as the weak coupling limit of a
random Schr\"odinger equation}, Comm. Pure Appl. Math. {\bf 53} (6), 667--735 (2000).
\bibitem{FeldKnoTru}
J. Feldman, H. Kn\"orrer, E. Trubowitz,
{\em A Two Dimensional Fermi Liquid,
Part 1: Overview},
Comm. Math. Phys, {\bf 247}, 1-47  (2004). 
\bibitem{HoLan}
T.G. Ho and L.J. Landau, {\em Fermi gas on a lattice in the van Hove limit}, J. Stat. Phys., {\bf 87},
821--845 (1997).
\bibitem{Hug}
N.M. Hugenholtz, {\em Derivation of the Boltzmann equation for a Fermi gas}, J. Stat. Phys., {\bf 32},
231--254 (1983).
\bibitem{Kl} 
A. Klein, 
{\em Absolutely continuous spectrum in the Anderson model on the Bethe lattice}, 
Math. Res. Lett. {\bf 1}, 399-407 (1994).
\bibitem{LukSpo}
J. Lukkarinen, H. Spohn, {\em Kinetic Limit for Wave Propagation in a Random Medium},
Arch. Ration. Mech. Anal. {\bf 183}, 93-162 (2007).
\bibitem{RodSchl} 
I. Rodnianski, W.  Schlag,   {\em Classical and
quantum scattering for a class of long range random potentials},
Int. Math. Res. Notices, {\bf 2003:5}, 243-300 (2003).
\bibitem{Sp}
H. Spohn, 
{\em Derivation of the transport equation for electrons moving through
random impurities}, J. Statist. Phys., 17, no. 6, 385-412 (1977).
\bibitem{Spo}
H. Spohn, {\em The phonon Boltzmann equation, properties and link to weakly anharmonic lattice dynamics},
J. Stat. Phys. {\bf 124},  no. 2-4, 1041--1104  (2006).
\end{thebibliography}
\end{document}